\documentclass[aps,prd,amsmath,floats,floatfix,onecolumn,superscriptaddress,nofootinbib]{revtex4-2}
\usepackage{graphicx}
\usepackage{dcolumn}
\usepackage{bm}

\usepackage[utf8]{inputenc}
\usepackage[T1]{fontenc}
\usepackage{hyperref}
\usepackage{fontawesome}
\usepackage{url}
\usepackage{amsfonts}
\usepackage{amsmath}
\usepackage{microtype}
\usepackage{xcolor}
\usepackage{natbib}
\usepackage{graphicx}
\usepackage{caption}
\usepackage{subcaption}
\usepackage{ragged2e}
\usepackage{booktabs}
\usepackage{placeins}
\usepackage[inline,shortlabels]{enumitem}
\usepackage{comment}

\usepackage{colortbl}
\definecolor{linkcolor}{rgb}{0.0,0.3,0.5}
\definecolor{darkred}{RGB}{175,0,0}
\definecolor{darkblue}{RGB}{14,0,185}
\definecolor{salmon}{RGB}{255,160,105}
\definecolor{redd}{RGB}{0,150,0}

\definecolor{darkteal}{RGB}{0,111,111}

\usepackage{tikz}
\usetikzlibrary{positioning,arrows.meta}

\providecommand{\aap}{Astron. Astrophys.}

\providecommand{\apj}{Astrophys. J.}

\providecommand{\mnras}{Mon. Not. R. Astron. Soc.}

\providecommand{\prd}{Phys. Rev. D}

\newcommand{\pcite}[1]{~\cite{#1}}
\newcommand{\req}[1]{Eq.~\eqref{#1}}
\newcommand{\rfig}[1]{Fig.~\ref{#1}}
\newcommand{\rsec}[1]{Sec.~\ref{#1}}
\newcommand{\rtab}[1]{Table~\ref{#1}}
\newcommand{\rapp}[1]{Appendix~\ref{#1}}

\newcommand{\bx}{\mathbf{x}}

\newcommand{\bk}{\mathbf{k}}
\newcommand{\bz}{\mathbf{z}}
\newcommand{\bS}{\mathbf{S}}
\newcommand{\btheta}{\boldsymbol{\theta}}

\newcommand{\Pdd}{P_{\delta\delta}}
\newcommand{\Pdm}{P_{\delta\Delta}}
\newcommand{\Pmm}{P_{\Delta\Delta}}
\newcommand{\kmax}{k_{\mathrm{max}}}
\newcommand{\kn}{k_{\mathrm{Nyq}}}
\newcommand{\dfilt}{\delta_{\mathrm f}}

\begin{document}

\title{Interpretable morphology-based marked statistics}

\newcommand\federicohome{
\affiliation{Department of Physics and Astronomy, Johns Hopkins University, Baltimore, MD 21218, USA}
}

\newcommand\alvisehome{
\affiliation{Dipartimento di Fisica Galileo Galilei, Universit\` a di Padova, I-35131 Padova, Italy}
\affiliation{INFN Sezione di Padova, I-35131 Padova, Italy}
\affiliation{INAF-Osservatorio Astronomico di Padova, Italy}
}

\newcommand\benhome{
\affiliation{Department of Physics and Astronomy, Johns Hopkins University, Baltimore, MD 21218, USA}
\affiliation{Department of Applied Mathematics and Statistics, Johns Hopkins University, Baltimore, MD 21218, USA}
}

\newcommand\michelehome{
\affiliation{Dipartimento di Fisica Galileo Galilei, Universit\` a di Padova, I-35131 Padova, Italy}
\affiliation{INFN Sezione di Padova, I-35131 Padova, Italy}
\affiliation{INAF-Osservatorio Astronomico di Padova, Italy}
}

\author{Federico Semenzato}
\email{semenzato@jhu.edu}
\federicohome

\author{Benjamin D. Wandelt}%
\benhome

\author{Michele Liguori}%
\michelehome

\author{Alvise Raccanelli}%
\alvisehome

\begin{abstract}
The shapes and environments of cosmic structures carry cosmological information beyond the matter power spectrum.
Recovering this information while retaining a physical understanding is a central challenge for cosmological inference.
We construct an interpretable, rotation-equivariant neural marking scheme that reweights the matter field, combining learned density responses with explicit measures of local morphology.
This approach enables direct inspection of the learned transformations, while retaining the advantages of a two-point function analysis.
We employ a contrastive learning framework to align the marked summaries with the respective cosmological parameters. 
In real-space matter simulations, the learned marked spectra recover information beyond the ordinary power spectrum, tightening marginalized Fisher constraints on~$\sigma_8$ and~$\Omega_m$ by factors of~$1.7\times$ and~$1.5\times$ relative to the power spectrum plus bispectrum at~$k_{\mathrm{max}}=0.3\,h\,\mathrm{Mpc}^{-1}$.
Without retraining, our mark also improves a neutrino-mass forecast by a factor of~$1.5\times$ relative to the power spectrum at the same scale cut.
Across the cosmological prior, our marking scheme also improve the prediction errors in both parameters by~$1.3\times$ over the best tested classical mark at~$k_{\mathrm{max}}=0.2\,h\,\mathrm{Mpc}^{-1}$.
A qualitative comparison with neural forecasts motivates the possibility that these different compressions approach a common information level.
Neural marked statistics thus offer a route to extracting non-Gaussian information from cosmological surveys, combining interpretable, physically motivated transformations with the practical advantages of two-point estimators.
\end{abstract}

\maketitle

\section{Introduction}
\label{sec:intro}
Gravitational evolution turns nearly Gaussian initial fluctuations into a network of halos, filaments, walls, and voids whose abundance, shapes, and density contrasts depend on cosmology\pcite{BonKofPog96,Hahn_2007,Whi14,1705.03021}.
Standard cosmological inference relies heavily on the two-point power spectrum, which is informationally complete for Gaussian fields and analytically tractable on large scales~\cite{Tegmark_1997,Percival_2004}. 
Yet, nonlinear structure formation couples Fourier modes and generates rich non-Gaussian morphology whose information content spills into higher-order correlators\pcite{Bernardeau:2001qr,Scoccimarro:1999ed,Sefusatti:2007fn,sabino_png,maldacena:2003,Bartolo:2004if,Dalal_2008,matarrese_effect_2008}. 
As ongoing galaxy surveys such as Euclid, DESI, SPHEREx, and Roman are mapping this structure over unprecedented volumes\pcite{desicollaboration2016desi,euclidcollaboration2024euclid,Dore:2014cca,spergel2015widefield}, we require summary statistics that can capture nonlinear structure while remaining practical to measure and interpret.

While higher-order correlation functions describe departures from Gaussianity explicitly, they represent a computational and analytical modeling challenge when moving to the nonlinear regime.
Moreover, the number of configurations grows rapidly with order, perturbative expansions diverge on small scales, and the Gaussian likelihood assumption breaks down.
Field-level inference, wavelet transforms, and neural summaries can recover information beyond low-order correlators\pcite{Leclercq:2021ctr,Nguyen:2024yth,Valogiannis_2022,imnn,Makinen:2021nly,SimBIG:2023ywd,Bairagi:2025ytq}, but depending on the inference procedure their chosen representation can be difficult to interpret.
Marked statistics are emerging as an alternative summary, designed to imprint a distortion on the field to make its two-point measurement sensitive to higher-order clustering\pcite{Massara:2020pli,Philcox:2020fqx,Marinucci:2024bdq}.
A mark~$M(\bx)$ reweights the density according to its local neighborhood, so the resulting power spectra mix higher-order correlations of the original field\pcite{Sheth_2005,Skibba_2006,WhiPad09,White:2016yhs,Philcox:2020fqx,Philcox:2020srd,Aviles_2019}.
The transformation can therefore bring additional information into a compact data vector measured with standard power-spectrum estimators.
Similarly, a related compression uses weighted skew-spectra: cross-spectra of the density with quadratic fields designed to retain bispectrum information for specified parameters\pcite{Schmittfull:2015skew,MoradinezhadDizgah:2020skew,Vanzan:2026ibo}.
This construction extends to the redshift-space galaxy bispectrum and has been applied to galaxy data\pcite{2021JCAP...03..020S,2024PhRvD.109j3528H}.

In a marked statistics, the choice of nonlinear transformation of the field determines which structures contribute to that information.
Density-dependent marks provide large forecast gains, especially for neutrino mass\pcite{Massara:2020pli}, and have been applied to modified gravity, biased tracers, and simulation-based inference\pcite{Valogiannis_2017,Armijo_2018,Ebina:2024zkv,2024SimbigMarks,Massara_2023}.
Recent studies optimize parametric mark families\pcite{Cowell:2024wyl,marinucci2026optimalmarkedstatisticseffective} or use the morphology of the local environment to weight the field\pcite{MartinBarandiaran:2026wwebs,Xiao:2026morphology}. 
In a previous work, we introduced rotation-equivariant neural marks with a contrastive objective that aligns marked spectra with cosmological parameters\pcite{Semenzato:2026stt}, creating an interpretable, flexible marking scheme that tightens parameter constraints across the prior.
Here we extend this program by organizing the mark around explicit descriptors of local morphology.

Morphology gives the mark freedom that density alone does not provide.
A single rotationally invariant amplitude of a tensor response measures its overall strength but does not retain the full distribution of its eigenvalues, which distinguish shapes of the local deformation with the same amplitude or trace.
In the Zel'dovich picture, collapse proceeds along the principal axes specified by the eigenvectors of the gravitational potential tidal tensor, producing sheets, filaments, and clusters\pcite{Zeldovich:1970,Hahn_2007,Bonnaire:2022web}. 
Counting eigenvalues of the potential tidal tensor above a threshold gives a corresponding four-environment classification\pcite{Hahn_2007,Forero-Romero_2009,1705.03021}.
Velocity-shear and density-Hessian classifiers apply related constructions to other tensors\pcite{Hoffman:2012ft,Aragon-Calvo:2004zco,Bond:2010ve,Cautun:2012ye}.
These descriptors provide a physical basis for marks that respond to density and local shape while remaining invariant under rotations.
We build a sequence of such marks, beginning with separate learned density responses in four fixed environments and then allowing those responses to overlap through soft curvature-dependent weights.
The filter, environment weights, and density responses remain explicit, so each ingredient can be inspected and tested by restricting or removing it.
The resulting Hessian Mark tightens the marginalized~$\sigma_8$ and~$\Omega_m$ constraints over the matter power spectrum by factors of~$6.5$ and~$3.0$ at~$\kmax=0.2\,h\,\mathrm{Mpc}^{-1}$, outperforms the power spectrum plus bispectrum at matched scale cuts, and retains a regression advantage over the tested classical marks across the prior.

The paper is organized as follows.
\rsec{sec:background} defines the marked spectra and the two mark constructions.
\rsec{sec:contrastive} describes their training and evaluation.
\rsec{sec:t_results} examines the learned transformations and their cosmological constraints, and \rsec{sec:t_conclusions} discusses the implications and next steps.
\rapp{app:t_ablations} tests the impact of the main design choices, and \rapp{app:round1_training} specifies the model and training settings.

\section{Methodology}
\label{sec:background}

A mark~$M(\bx)$ is a local functional of the density contrast~$\delta(\bx)$ in a neighborhood of~$\bx$ that reweights the field before the two-point measurement.
The marked field is then
\begin{equation}
  \Delta(\bx) = M(\bx)\,\bigl[1+\delta(\bx)\bigr] - \bigl\langle M(1+\delta)\bigr\rangle\,,
  \label{eq:marked_field}
\end{equation}
where the angle brackets denote the average over the field.
We summarize the marked and unmarked fields by their two auto-spectra and their cross-spectrum,
\begin{equation}
  \bS(k)=\left\{\Pdd(k),\ \Pdm(k),\ \Pmm(k)\right\}\,,
  \label{eq:spectra_set}
\end{equation}
with the auto- and cross-spectra of the marked and unmarked fields defined as
\begin{equation}
\bigl\langle a(\bk)\,b^\ast(\bk')\bigr\rangle = (2\pi)^3\,\delta_D(\bk-\bk')\,P_{ab}(k)\,,
\qquad a,b\in\{\delta,\Delta\}\,.
\label{eq:t_spectra_def}
\end{equation}
Here~$\delta_D$ is the Dirac delta function, and we estimate the power spectra in Fourier shells of width~$k_F=2\pi/L$, where~$L$ is the side length of the simulation box.
For comparison with a higher-order statistic, we concatenate the unmarked power spectrum with the matter bispectrum~$B$, defined by 
\begin{equation}
  \langle \delta(\bk_1)\delta(\bk_2)\delta(\bk_3)\rangle=(2\pi)^3\delta_D(\bk_1+\bk_2+\bk_3)B(k_1,k_2,k_3)\,.
  \label{eq:spectra_set_bispec}
\end{equation}
We denote this joint data vector by P+B and include its full cross-covariance.
We retain every available bispectrum bin whose three side-bin centers satisfy~$k_i\leq\kmax$.
The gain over the power spectrum comes from the nonlinearity of the mark.
Expanding the mark in powers of~$\delta$ expresses~$\Pdm$ and~$\Pmm$ in terms of higher-order density correlations.
A well-chosen mark thus folds bispectrum- and trispectrum-level information into a measurement that remains a power spectrum\pcite{Philcox:2020fqx,Philcox:2020srd,Aviles_2019,Marinucci:2024bdq,Ebina:2026qzf}.
The mark can moreover bring sensitivity to nonlinear structure into wavelengths where the unmarked density is approximately linear\pcite{Philcox:2020fqx,Marinucci:2024bdq,marinucci2026optimalmarkedstatisticseffective}.
The Fisher information of the full set~$\bS$ is invariant under parameter-independent affine maps~$M\to aM+c$ with~$a\neq0$\pcite{Cowell:2024wyl}.
The induced map~$\Delta\to a\Delta+c\delta$ is linear and invertible, as is its action on the three spectra and their covariance, therefore the mark shape is the only relevant feature for the information content of~$\bS$.

The so-called classical mark is a function of one scalar, the density contrast~$\delta_R$ smoothed on a scale~$R$\pcite{White:2016yhs,Massara:2020pli},
\begin{equation}
  M(\bx)=\left(\frac{1+\delta_s}{1+\delta_s+\delta_R(\bx)}\right)^p\,,
  \label{eq:classical_mark}
\end{equation}
where, for~$\delta_s>0$, a positive (negative) exponent~$p$ upweights (downweights) underdense regions.
We adopt this configuration as the reference \emph{classical marks}, spanning a range of~$\delta_s$,~$R$ and~$p$ to compare with the learned neural marks.

\subsection{Tidal field and morphology of the cosmic web}
\label{sec:class_morpho}

\begin{figure}[t]
  \centering
  \includegraphics[width=\textwidth]{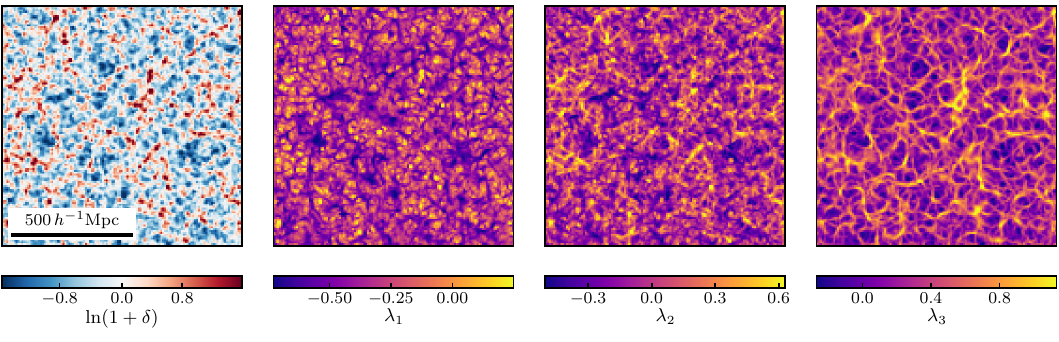}
  \caption{\justifying The ordered eigenvalues distinguish local deformation shapes with the same tidal-tensor trace.
From left to right: the log-density~$g=\ln(1+\delta)$ and the ordered eigenvalues~$\lambda_1\le\lambda_2\le\lambda_3$ of the potential tidal tensor sourced by~$g$, on the same fiducial slice.
Their signs define the void, sheet, filament, and cluster classes of \req{eq:t_classification}.
}
  \label{fig:t_pot_eigs}
\end{figure}

\label{sec:theory_tweb}
In the Zel'dovich picture, a mass element contracts sequentially along the principal axes of the local deformation, turning a smooth perturbation first into a sheet, then a filament, then a cluster\pcite{Whi14}.
For a rescaled gravitational potential~$\Phi$ satisfying~$\nabla^2\Phi=\delta$, the symmetric tidal tensor describes the local deformation. In real and Fourier space, 
\begin{equation}
T_{ij}(\bx) = \partial_i\partial_j\,\Phi(\bx)\,,
\qquad
\widetilde{T}_{ij}(\bk) = \frac{k_i k_j}{k^2}\,\widetilde{\delta}(\bk)\,.
\label{eq:t_tensor}
\end{equation}
We order the eigenvalues of~$T_{ij}$ as~$\lambda_1\le\lambda_2\le\lambda_3$.
Their sum is the density contrast,~$\lambda_1+\lambda_2+\lambda_3=\delta$, while their relative values describe how the deformation is shared among the three principal axes.
Functions of the eigenvalues retain this shape information and discard the axis orientations carried by the eigenvectors.
Linearizing the relative acceleration in a frozen potential leads to~$\ddot{x}_i=-T_{ij}x_j$.
In the eigenframe, a positive eigenvalue hence implies an inward acceleration along its axis, motivating a classification by the number of compressive directions\pcite{Hahn_2007}.
Counting collapsing directions therefore assigns every point to one of four cosmic-web environments\pcite{Hahn_2007,Forero-Romero_2009}
\begin{equation}
n_{\mathrm{coll}}(\bx) = \sum_{i=1}^{3}\Theta(\lambda_i - \lambda_{\mathrm{th}})
\;\in\;\{0,1,2,3\}
\;\longrightarrow\;
\{\text{void, sheet, filament, cluster}\}\,,
\label{eq:t_classification}
\end{equation}
with~$\Theta$ the Heaviside function. 
The threshold~$\lambda_{\mathrm{th}}$ controls which directions count as collapsing and thus changes the assigned environments\pcite{Forero-Romero_2009}.
We fix~$\lambda_{\mathrm{th}}=0$ throughout, so that the boundaries between the four classes are set by the signs of the eigenvalues.
The eigenstructure of~$T_{ij}$ hence separates populations through anisotropic, higher-order combinations of the field that a function of~$\delta_R$ cannot represent.
Since the eigenvalues are nonlinear functions of the second derivatives of the potential, a nonlinear function of them can capture higher-order correlators of the original field, motivating the use of local morphology to drive the design of a marking scheme.

\begin{figure}[t]
  \centering
  \begin{minipage}[t]{0.49\textwidth}
    \centering
    \includegraphics[width=\linewidth]{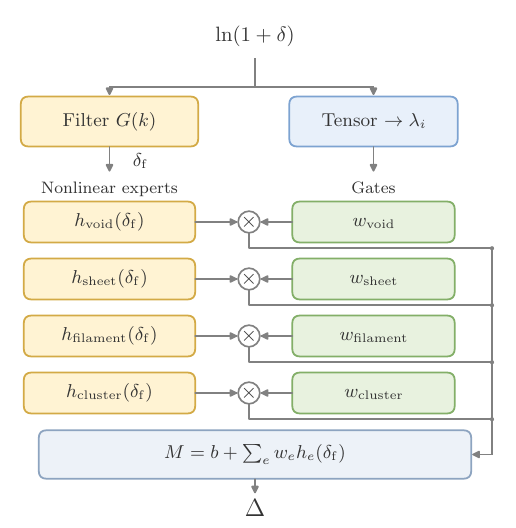}\par\vspace{3pt}
    {\small (a) Marking network}
  \end{minipage}\hfill
  \begin{minipage}[t]{0.49\textwidth}
    \centering
    \includegraphics[width=\linewidth]{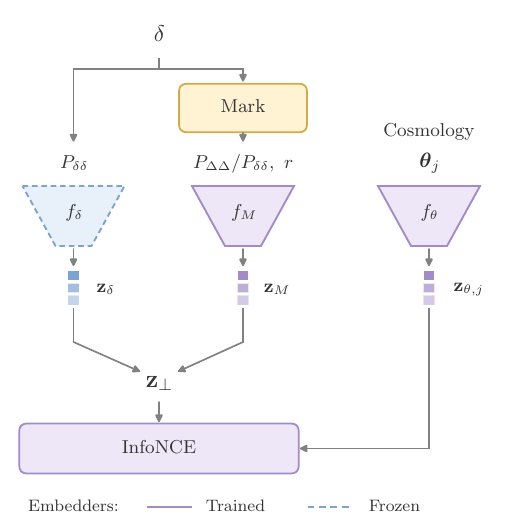}\par\vspace{3pt}
    {\small (b) Contrastive framework}
  \end{minipage}
  \caption{\justifying Local density responses are mixed by geometry and trained to distinguish cosmologies.
\emph{(a) Marking network}: a shared radial filter~$G(k)$ feeds four nonlinear density experts~$h_e$, whose outputs are weighted by gates~$w_e$ computed from tensor eigenvalues (hard for the Web Mark, soft for the Hessian Mark). Summing these products with the offset~$b$ gives the mark~$M$ of \req{eq:t_combiner}; reweighting the density and subtracting the mean provides~$\Delta$ of \req{eq:marked_field}.
\emph{(b) Contrastive framework}, following Ref.~\cite{Semenzato:2026stt}: a trained embedder maps the marked amplitude and cross-correlation features of \req{eq:t_summary_features} to~$\bz_M$; removing its component along the frozen unmarked embedding~$\bz_\delta$ gives~$\bz_\perp$ (\req{eq:t_ortho}), which an InfoNCE loss matches to the embedding~$\bz_\theta$ of the generating cosmology from a linear parameter projector.}
  \label{fig:prd_marker_scheme}
  \label{fig:prd_contrastive_scheme}
\end{figure}

At each position, the marking scheme separates the response to the surrounding neighborhood density (supplied from the filtered log-density) from the response to the local field shape (provided by the tensor eigenvalues).
Each learned density response (an expert) specifies a weighting as a function of neighborhood density, while the gates determine which responses contribute at a given position.
This mixture of experts allows two cells with the same neighborhood density to receive different marks when their local geometry differs.
\rfig{fig:prd_marker_scheme} summarizes the construction.

As a first tidal mark variant (labeled \textit{Web Mark}), we evaluate the tensor of \req{eq:t_tensor} on the log-density~$g=\ln(1+\delta)$, choosing the logarithm compression to control the dynamic range of collapsed cores\pcite{NeySzaSza09,Neyrinck_2011}.
The periodic potential is sourced by the centered log-density,~$\nabla^2\Phi_g=g-\langle g\rangle$, with its zero Fourier mode set to zero.
The inverse Laplacian, computed with a fast Fourier transform (FFT), acts only on nonzero modes, so~$\operatorname{tr}T=g-\langle g\rangle$.
Its eigenvalues retain the geometric construction of the tidal classifier, but the potential sourced by~$g$ is not the gravitational potential sourced by~$\delta$.

A learned radial filter~$G(k)$, shared by all environments, maps the log-density to the filtered field~$\dfilt=\mathcal{F}^{-1}[G(k)\,\widetilde{g}(\bk)]$.
The profile~$G(k)$ is parameterized by learnable weights on a uniform grid of~$32$ knots in~$|\bk|$, passed through a sigmoid, linearly interpolated onto the Fourier grid, and divided by its value at~$k=0$, so that~$G(0)=1$.
The profile is initialized as a Gaussian low-pass.
We assign every cell~$\bx$ to one environment~$e$ by the count rule of \req{eq:t_classification}, written as a gate~$w_e(\bx)\in\{0,1\}$.
Each environment has an independent learned density response, or expert,~$h_e(\dfilt)$.
The gates select the appropriate response in each cell, so that the mark is a sum of the four environment-specific contributions,
\begin{equation}
    M(\bx) = b + \sum_{e}\, w_e(\bx)\,h_e\!\bigl(\dfilt(\bx)\bigr)\,,
\label{eq:t_combiner}
\end{equation}
where the sum runs over~$e\in\{\mathrm{void},\mathrm{sheet},\mathrm{filament},\mathrm{cluster}\}$, with the functions~$h_e$ zero-initialized and the trainable offset~$b$ initialized at~$1$, so training starts from~$M\equiv1$.

We can further extend the freedom of the strict morphology classifiers.
In this generalized version (labeled \textit{Hessian Mark}), we replace the hard classes with overlapping weights based on local curvature.
Its gates read the sign-reversed Hessian of the log-density, evaluated in Fourier space,
\begin{equation}
H_{ij}(\bx) = -\,\partial_i\partial_j\,g(\bx)\,,
\qquad
\widetilde{H}_{ij}(\bk) = k_i k_j\,\widetilde{g}(\bk)\,,
\label{eq:t_hessian}
\end{equation}
where the sign makes all three eigenvalues positive at non-degenerate log-density maxima.
The Fourier implementation and eigenvalue normalization are specified in \rapp{app:round1_training}.
Relative to the potential tidal tensor sourced by~$g$, the Hessian has an extra Fourier-space factor of~$k^2$, i.e.,~$\widetilde{H}_{ij}=k^2\widetilde{T}_{ij}$.
Its eigenvalues therefore emphasize cell-scale curvature.
The Hessian describes whether the log-density bends upward or downward along each principal direction: a trough is convex along all three, a peak is concave along all three, and saddle-like structures have mixed signs.
The eigenvalues provide local shape descriptors; identifying an extended wall or filament would also require spatial coherence across cells.
Together with~$\dfilt$, the eigenvalues let the density response vary with the sharpness and shape of structure inside the surrounding neighborhood.

We extract the ordered eigenvalues~$\lambda_1\!\le\!\lambda_2\!\le\!\lambda_3$ in closed form for a symmetric~$3\times3$ matrix.
We replace the hard count in \req{eq:t_classification} with a differentiable partition of unity. 
With a temperature~$\tau$ and monotone thresholds~$\lambda_{\mathrm{th},1}\!\ge\!\lambda_{\mathrm{th},2}\!\ge\!\lambda_{\mathrm{th},3}$, the sigmoid responses~$p_i=\sigma\!\bigl((\lambda_i/s-\lambda_{\mathrm{th},i})/\tau\bigr)$ soften the eigenvalue thresholds, with~$\sigma(u)=(1+e^{-u})^{-1}$.
The constant~$s$ is the standard deviation of the three eigenvalues pooled over all cells of eight fiducial boxes, computed once and used unchanged for every box.
The thresholds and the temperature are therefore in units of~$s$: on the eigenvalue itself the threshold is~$s\,\lambda_{\mathrm{th},i}$ and the temperature~$s\,\tau$.
The four environment weights can then be expressed as\pcite{Rossello:2026fuzzy} 
\begin{equation}
w_{\mathrm{void}} = 1-p_3\,,\quad
w_{\mathrm{sheet}} = p_3-p_2\,,\quad
w_{\mathrm{filament}} = p_2-p_1\,,\quad
w_{\mathrm{cluster}} = p_1\,.
\label{eq:t_gates}
\end{equation}
The weights sum to one and are non-negative because the eigenvalue and threshold orderings guarantee~$p_1\!\le\!p_2\!\le\!p_3$.
As in the Web Mark, we fix the thresholds at zero and we choose the fixed temperature~$\tau=0.6$ after calibration.
These gates smoothly extend the hard eigenvalue-sign classification while preserving a partition of unity. They recover hard assignments as~$\tau\to0$ away from zero eigenvalues and blend the environment-specific experts near boundaries at finite~$\tau$.
The density experts use the same filtered log-density~$\dfilt$ as before, while the gates read the unfiltered Hessian.
Using the same scheme as in \req{eq:t_combiner}, we can then combine the expert responses with the soft weights of \req{eq:t_gates}.
Each contribution is explicit, allowing us to inspect both its density dependence and its association with the four environment labels.

\section{Training and evaluation}
\label{sec:contrastive}
\subsection{Contrastive framework}
We train the marks with the contrastive framework of Ref.~\cite{Semenzato:2026stt}. 
For each simulated field, the contrastive task identifies the respective parameters among alternative cosmologies, so a useful mark must retain parameter-dependent structure despite realization-to-realization fluctuations.
The unmarked spectrum provides a reference representation against which the marked contribution is trained (see \rfig{fig:prd_contrastive_scheme}).

From the spectra of \req{eq:spectra_set}, measured at~$k\le0.3\,h\,\mathrm{Mpc}^{-1}$, the networks receive the unmarked amplitude and the marked amplitude and cross-correlation channels
\begin{equation}
\bS_\delta = \log_{10}\Pdd(k)\,,\qquad
\bS_M(k) = \left[\,\log_{10}\frac{\Pmm(k)}{\Pdd(k)},\ r(k)\,\right]\,.
\label{eq:t_summary_features}
\end{equation}
with~$r(k)=\Pdm(k)/\sqrt{\Pdd(k)\Pmm(k)}$ the cross-correlation coefficient of the marked and unmarked fields.

Summaries ($\bS$) and parameters ($\btheta$) are then projected onto unit vectors of a shared~$16$-dimensional space through dedicated embedders~$f_\alpha$ ($\bz_\alpha=f_\alpha(\bS_\alpha)$, with~$\alpha\in\{\delta,M\}$) and a linear parameter projector ($\bz_\theta=f_\theta(\btheta)$).
Training proceeds in two stages.
The first is a pretraining stage that aligns the unmarked embedder and parameter projector to avoid a collapsed unmarked representation that would make the complementarity constraint meaningless.
In the second, we freeze the unmarked embedder and train the mark, marked embedder, and parameter projector.
We subtract the projection of the marked embedding onto the unmarked one and renormalize the residual,
\begin{equation}
\bz_\perp = \frac{\bz_M - (\bz_M\!\cdot\!\bz_\delta)\,\bz_\delta}{\bigl\|\bz_M - (\bz_M\!\cdot\!\bz_\delta)\,\bz_\delta\bigr\|}\,,
\label{eq:t_ortho}
\end{equation}
so the objective favors embedding directions orthogonal to those of the unmarked summary.
This constraint encourages complementary representations, but does not guarantee statistical independence or additional Fisher information; the forecasts and regression tests assess the resulting gain.
The orthogonalized embedding~$\bz_\perp$ is then aligned to the parameter embedding~$\bz_\theta$ through an InfoNCE loss\pcite{Oord_2018}
\begin{equation}
\mathcal{L}_i = -\log\frac{\exp \left(\operatorname{sim}(\bz_{\perp,i},\bz_{\theta,i})\right)}{\sum_{j\in\mathcal{C}_i}\exp \left(\operatorname{sim}(\bz_{\perp,i},\bz_{\theta,j})\right)}\,,
\label{eq:t_infonce}
\end{equation}
where the candidate set~$\mathcal{C}_i$ contains the true parameters of simulation~$i$ and negative examples drawn from other training simulations and from global and local samples of the prior\pcite{Semenzato:2026stt}.
The similarity is the cosine between the unit vectors at a fixed temperature~$\tau_c=0.05$
\begin{equation}
\operatorname{sim}(\bz_\perp,\bz_\theta) = \bz_\perp\!\cdot\!\bz_\theta/\tau_c\,,
\label{eq:t_similarity}
\end{equation}
and both training stages use this similarity.
After training, we fix the mark and measure~$\{\Pdd,\Pdm,\Pmm\}$ for inference.
The model architectures and training hyperparameters are given in \rapp{app:round1_training}, in \rtab{tab:round1_architecture} and \rtab{tab:round1_optimization}, respectively.

\subsection{Evaluating information gain}
\label{sec:t_evaluation}

We evaluate the gain over the matter power spectrum through a Fisher forecast at the fiducial cosmology and parameter regression across the training prior. 
For a data vector~$\bm{\mu}(\btheta)$ (i.e., the mean of a summary such as~$\bS$ of \req{eq:spectra_set} over the bins with~$k\le\kmax$),  with covariance~$C$, from a Gaussian likelihood with parameter-independent covariance we can define the Fisher matrix\pcite{Jungman95,vogeley:fisher,Tegmark:1996bz,Tegmark_1997}
\begin{equation}
  F_{\alpha\beta}=\frac{\partial\bm{\mu}^{\top}}{\partial\theta_\alpha}\,C^{-1}\,\frac{\partial\bm{\mu}}{\partial\theta_\beta}\,,
  \label{eq:t_fisher}
\end{equation}
evaluated at the fiducial cosmology for the five parameters~$\btheta=\{\Omega_m,\Omega_b,h,n_s,\sigma_8\}$.
Within the assumed likelihood, the marginalized Fisher error~$\sigma_\alpha=[(F^{-1})_{\alpha\alpha}]^{1/2}$ sets the Cram\'er--Rao lower bound on the standard deviation of an unbiased estimator of~$\theta_\alpha$.
We marginalize every error over the other four parameters unless otherwise specified.
The derivatives are central differences on the Quijote derivative suite, with~$500$ realizations per side\pcite{Villaescusa-Navarro:2019bje}.
The covariance is computed from~$8{,}000$ fiducial realizations, with the Hartlap correction applied to its inverse\pcite{Hartlap:2007}.
We summarize performance relative to a reference summary~$\bS_{\rm ref}$ by the error ratio
\begin{equation}
  \mathcal{I}_\alpha(\bS_{\rm ref};\kmax)=\frac{\sigma_\alpha\bigl(\bS_{\rm ref};\kmax\bigr)}{\sigma_\alpha\bigl(\bS;\kmax\bigr)}\,,
  \label{eq:t_improvement}
\end{equation}
where numerator and denominator use the same parameter set and scale cut.
Unless another reference is specified,~$\bS_{\rm ref}=\Pdd$ and we omit this argument.
For comparisons with power spectrum and bispectrum (P+B) or a classical marked summary~$\bS_{\rm C}$, we write~$\mathcal{I}_\alpha(\mathrm{P+B})$ or~$\mathcal{I}_\alpha(\bS_{\rm C})$, respectively, suppressing~$\kmax$ when it is clear.
In every case,~$\mathcal{I}_\alpha>1$ means that the tested summary has a smaller error than the stated reference.
To test performance across the prior, we train a multilayer perceptron to predict the five parameters from \req{eq:t_summary_features}, using a subset of simulations that is disjoint from those used to train the marks and covers the same prior range.
Features and targets are standardized using the training set only; the five network-initialization seeds are paired across summaries.
For this evaluation we compare the root mean squared error~$\mathrm{RMSE}_\alpha(\bS)=\sqrt{\langle(\hat\theta_\alpha-\theta_\alpha)^2\rangle_{\rm test}}$.
We denote the corresponding reference-to-learned RMSE ratio by~$\mathcal{I}_\alpha^{\rm reg}(\bS_{\rm C})=\mathrm{RMSE}_\alpha(\bS_{\rm C})/\mathrm{RMSE}_\alpha(\bS)$.
We use the same network as Ref.~\cite{Semenzato:2026stt} and train it with the log-MSE loss~$\mathcal{L}=\sum_\alpha\ln(\mathrm{MSE}_\alpha+\epsilon)$, with~$\epsilon=10^{-6}$ and MSE in physical parameter units.
The derivative of this loss weights each parameter by~$1/(\mathrm{MSE}_\alpha+\epsilon)$, reducing sensitivity to differences in parameter scales while bounding the weight for small errors.

\section{Results}
\label{sec:t_results}
The decomposition into gates, density responses, and their sum makes the learned transformation directly inspectable.
We first explore these components, then test whether their added structure improves cosmological constraints.
In \rapp{app:t_ablations}, we test how logarithmic compression, learned density responses, curvature gates, and additional components affect the final marginalized constraints by removing or restricting these ingredients.

For training, we use the matter density contrast at~$z=0$ from the Quijote simulation suite\pcite{Villaescusa-Navarro:2019bje}, assigned to~$128^3$ grids in~$(1\,h^{-1}\mathrm{Gpc})^3$ boxes with the piecewise-cubic-spline scheme\pcite{Hand_2018}.
The cell size is therefore~$7.8\,h^{-1}\mathrm{Mpc}$ and the Nyquist wavenumber~$\kn=0.40\,h\,\mathrm{Mpc}^{-1}$.
The marks train on a separate suite of~$\sim5{,}000$ simulations of the Big Sobol Sequence (BSQ) \pcite{Bairagi:2025sux}, with~$15\%$ held out for validation.
Its five parameters follow a Sobol sequence over~$\Omega_m\in[0.1,0.5]$,~$\Omega_b\in[0.02,0.08]$,~$h\in[0.5,0.9]$,~$n_s\in[0.8,1.2]$, and~$\sigma_8\in[0.6,1.0]$.
We train each mark with three independent random initialization seeds.
We focus the quantitative tests on the Hessian Mark, which provides tighter fiducial constraints than the Web Mark; we retain the latter as a comparison in \rfig{fig:t_constraints}.
Unless stated otherwise, results and maps refer to the Hessian Mark with~$\tau=0.6$ from the training with the largest marginalized~$\sigma_8$ improvement at~$\kmax=0.2\,h\,\mathrm{Mpc}^{-1}$.
The code for the model and postprocessing is available at \href{https://github.com/fsemenzato/Interpretable-Equivariant-Marks.git}{\faGithub}.

\subsection{Interpreting the learned marks}
\label{sec:t_introspection}

\begin{figure*}[t]
  \centering
  \includegraphics[width=\textwidth]{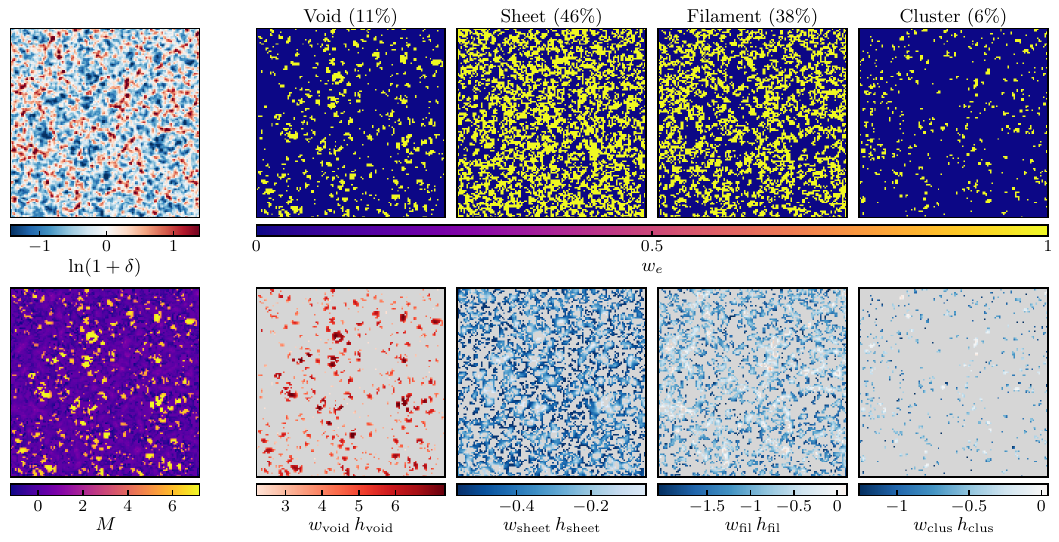}
  \caption{\justifying  The Web Mark mainly distinguishes void regions from the other environments, with only a mild density response within each class.
The left column shows the log-density (top) and full mark~$M$ (bottom) on the same fiducial realization.
The four columns on the right show the hard masks~$w_e$ (top, with volume fractions) and their contributions~$w_eh_e$ (bottom); gray cells lie outside the corresponding mask.
The four contributions and the constant~$b$ sum to the full mark of \req{eq:t_combiner}.}
  \label{fig:t_web_mark}
\end{figure*}

\subsubsection{Web Mark}

\rfig{fig:t_web_mark} shows the trained Web Mark applied to one slice at the fiducial cosmology.
Voids, sheets, filaments, and clusters fill approximately~$11\%$,~$46\%$,~$38\%$, and~$6\%$ of the volume, respectively (rounded independently).
With voxels of~$7.8\,h^{-1}\mathrm{Mpc}$, walls and filaments are at most a few cells thick, and the tidal tensor eigenvalues vary from one cell to the next about as much as the log-density itself.
Each label combines the signs of three such eigenvalues, so neighboring cells often fall into different environments.
The trace equals the log-density relative to its box mean,~$g-\langle g\rangle$, linking the sign-count classification to this centered source.
Within each environment the expert responses are small and smooth.
The void mask still follows the extended low-density regions, while cluster cells lie near dense concentrations.
The classification therefore retains recognizable morphology, although the grid resolves walls and filaments poorly.
The trained Web Mark uses this partition mainly as a void indicator with a mild density response within each class.

\begin{figure*}[t]
  \centering
  \includegraphics[width=\textwidth]{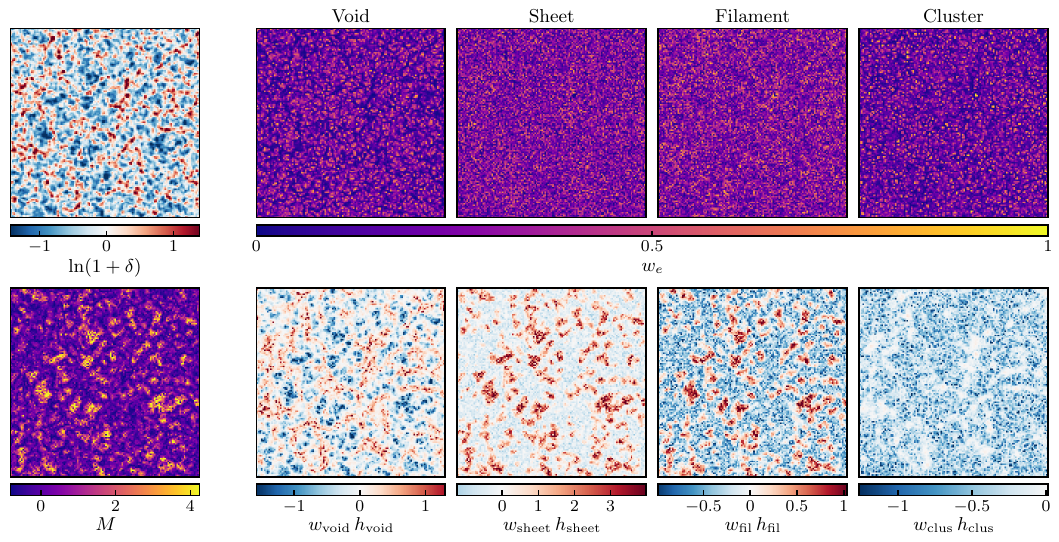}
  \caption{\justifying The Hessian Mark combines overlapping environment weights with learned density responses, while retaining visible void and peak structure.
The layout and slice match \rfig{fig:t_web_mark}: full log-density and mark on the left, soft gates~$w_e$ above their contributions~$w_eh_e$ on the right.
The environment names label the gate construction, while each cell can contribute to several terms.}
  \label{fig:t_hm_gates}
\end{figure*}

\subsubsection{Hessian Mark}

The decomposition in \rfig{fig:t_hm_gates} shows how these curvature weights enter the final mark.
The gates (top row) retain a visible correspondence with morphological features: void-like weights favor low-density regions, while cluster-like weights favor peaks.
The gates are broad, overlapping weights rather than mutually exclusive environment labels, with the largest weight exceeding~$0.8$ in only~$1\%$ of cells.
The overlap does not remove their connection to morphology: void and cluster regions remain clearly distinguishable, although the~$7.8\,h^{-1}\mathrm{Mpc}$ cells limit the spatial detail of walls and filaments.
The extreme eigenvalues connect these weights to the local curvature.
If even the largest eigenvalue~$\lambda_3$ is negative, the log-density is locally convex along all three principal axes; if even the smallest eigenvalue~$\lambda_1$ is positive, it is locally concave along all three.
The void and cluster gates in \req{eq:t_gates} soften these two conditions, favoring trough-like and peak-like curvature, respectively.
The gates can therefore assign different expert mixtures to cells with similar filtered neighborhood density but different local curvature.

\begin{figure*}[t]
  \centering
  \includegraphics[width=\textwidth]{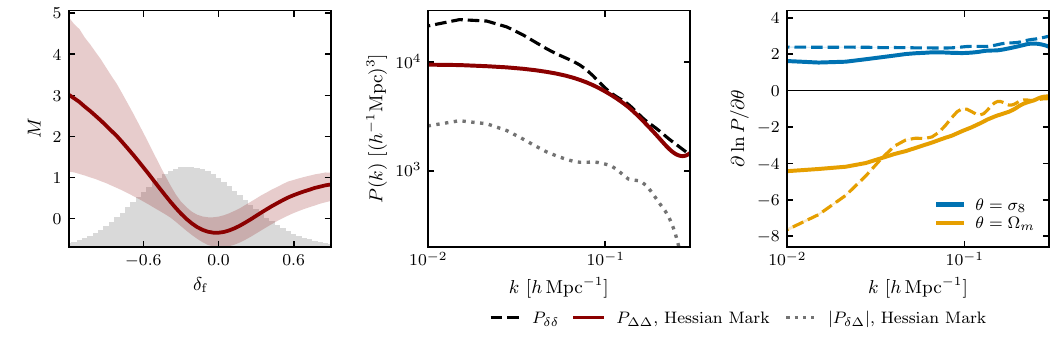}
  \caption{\justifying The Hessian Mark weights underdense neighborhoods strongly, and its spectra respond to cosmology differently from the matter power spectrum, which lets the joint summary constrain parameter combinations that~$\Pdd$ alone measures poorly.
\emph{Left}: mean~$M$ as a function of the filtered log-density~$\dfilt$ over eight fiducial boxes; the shaded band spans the~$16$th to~$84$th percentiles of cells in each bin, and the gray histogram shows their relative abundance.
\emph{Center}: fiducial auto-spectra and the absolute cross-spectrum.
\emph{Right}: logarithmic derivatives of~$\Pdd$ (dashed) and~$\Pmm$ (solid) with respect to~$\sigma_8$ and~$\Omega_m$.
}
  \label{fig:t_hm_slices}
  \label{fig:t_hm_spectra}
\end{figure*}

\rfig{fig:t_hm_slices} (left) summarizes the final mark as a function of the filtered log-density~$\dfilt$.
Since the mark is not a unique function of the density value, the shaded band spans the~$16$th to~$84$th percentiles of cells in each density bin to visualize the cell-to-cell variation.
The gray histogram shows the relative volume in each density bin, so the strength of the mean response can be read alongside the abundance of the corresponding neighborhoods.
The mean response is lowest near~$\dfilt=0$ and rises on both sides, steeply toward underdense neighborhoods and more gently toward overdense ones.
At fixed~$\dfilt$, this curve averages the products of each expert response and its conditional gate weight, with both varying with neighborhood density.
The profile learned from the contrastive objective is enhancing the environment contrast, while the absolute mark amplitude is not a relevant feature, since affine changes of~$M$ preserve the information in the full summary (see \rsec{sec:background}).
The spread reflects different local Hessian eigenvalues, and thus different gate weights, among cells with similar neighborhood density.
The profile therefore describes the mean density weighting without specifying the complete transformation.
The product~$w_e(\bx)h_e[\dfilt(\bx)]$ couples the local curvature to the neighborhood density before~$M$ multiplies~$1+\delta$.
The two-point measurement consequently responds both to the mixture of local structures present in a realization and to where those structures lie relative to the density field.
Neither the volume occupied by a gate nor the visual prominence of one contribution alone measures its cosmological information: the forecast depends on the joint responses and covariance of all three spectra.

In \rapp{app:round1_gate_interventions}, we distinguish the role of the gates by replacing them with their box averages in each realization.
This test removes the spatial dependence of the weights while retaining a box-dependent mixture of the same density experts.
The~$\sigma_8$ improvement remains close to the trained value, whereas using the same fixed mixture in every realization substantially weakens it.
Much of the amplitude sensitivity therefore enters through changes in the distribution of cell-scale curvature.
For~$\Omega_m$ and~$n_s$, removing the spatial pattern weakens the improvement more strongly, indicating that the registration of curvature with density also matters.
The results therefore support a role for morphology through both local shape-dependent weighting and the abundance of curvature responses, without identifying the full gain with the resolved arrangement of walls and filaments.

\subsection{Constraints at the fiducial cosmology and across the parameter space}
\label{sec:t_constraints}

We now quantify how much cosmological information our neural marks add to the power spectrum, using~$\mathcal{I}_\alpha$ of \req{eq:t_improvement}.
We use the three spectra~$\{\Pdd,\Pdm,\Pmm\}$ of \req{eq:spectra_set}, beginning with the fiducial forecast and its extension to neutrino mass before testing regression across the prior.
At the fiducial cosmology, we evaluate both the Web Mark and the Hessian Mark, varying~$\{\Omega_m,\Omega_b,h,n_s,\sigma_8\}$ and marginalizing over the other four parameters when quoting each error.
We estimate the joint covariance of each three-spectrum summary from~$8{,}000$ fiducial realizations and the parameter derivatives from~$500$ realizations per side, with the Hartlap correction applied to the inverse covariance.
For the direct comparison with P+B, we use the same~$8{,}000$ covariance realizations for all summaries.
Unless explicitly comparing different cuts below, all measured spectra and bispectrum triangle side-bin centers use the same~$\kmax$.
\rfig{fig:t_hm_spectra} (center and right) shows the spectra and their responses to~$\sigma_8$ and~$\Omega_m$.
On large scales the matter power spectrum responds to~$\sigma_8$ as linear theory predicts,~$\partial\ln\Pdd/\partial\sigma_8\simeq2/\sigma_8\approx2.4$, since~$\Pdd\propto\sigma_8^2$.
Both logarithmic~$\Omega_m$ responses remain negative throughout the plotted range. Their magnitudes are largest at small~$k$ and decrease toward higher~$k$, with expected oscillatory features in the unmarked response.
The marked spectra have different responses, and the resulting joint constraints depend on these derivatives and the full covariance, including the cross-spectrum.
Their joint measurement allows us to constrain parameter combinations that~$\Pdd$ alone measures poorly.
Moreover, an affine transformation of the mark~$M\to aM+b$, with parameter-independent constants~$a$ and~$b$, imply~$\Delta\to a\Delta+b\delta$ by \req{eq:marked_field}.
The spectra therefore transform as~$\Pdm\to a\Pdm+b\Pdd$ and~$\Pmm\to a^2\Pmm+2ab\Pdm+b^2\Pdd$, while~$\Pdd$ is unchanged.
For~$a\ne0$, this mixing can change the individual marked responses but leaves the Fisher information of the joint three-spectrum summary unchanged when its full covariance is transformed consistently \pcite{Cowell:2024wyl,marinucci2026optimalmarkedstatisticseffective}.
The response of~$\Pmm$ alone thus depends on the plotted convention even though the information of their full set does not.

\begin{figure*}[t]
  \centering  \includegraphics[width=\textwidth]{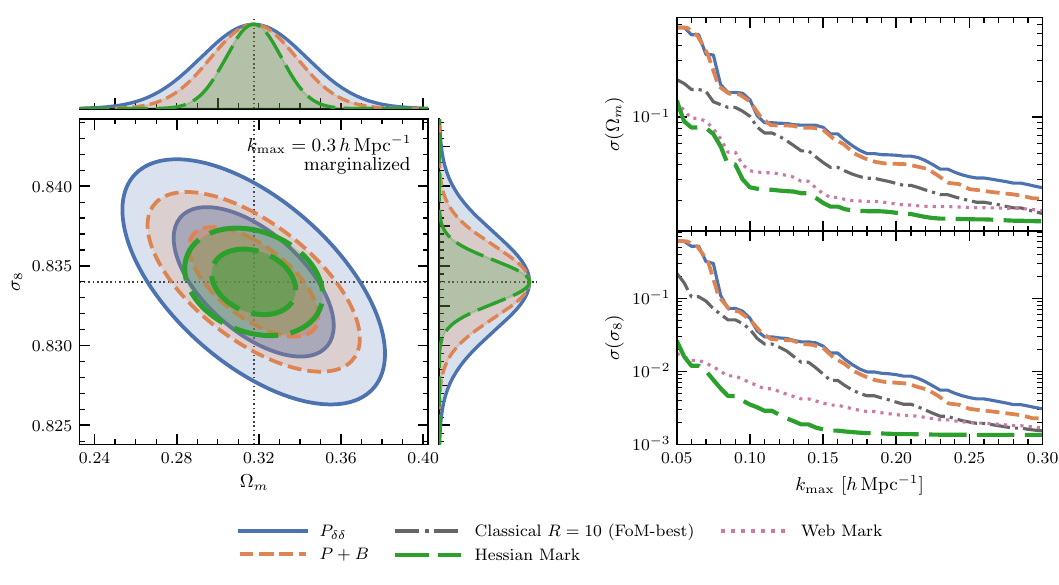}
  \caption{\justifying  The Hessian Mark improves the marginalized constraints beyond P+B at matched scale cuts, with its largest gain over~$\Pdd$ at low~$\kmax$, where its~$\sigma_8$ error has already nearly converged. 
\emph{Left}:~$68\%$ and~$95\%$ Fisher regions in~$(\Omega_m,\sigma_8)$ at~$\kmax=0.3\,h\,\mathrm{Mpc}^{-1}$, marginalized over~$\Omega_b$,~$h$, and~$n_s$.
\emph{Right}: marginalized errors versus~$\kmax$. The classical mark maximizes~$\sqrt{\det F}$ at~$\kmax=0.2\,h\,\mathrm{Mpc}^{-1}$ among the~$R=10\,h^{-1}\mathrm{Mpc}$ variants explored in \rfig{fig:t_family}.
Each marked summary uses~$\{\Pdd,\Pdm,\Pmm\}$.}
\label{fig:t_constraints}
\end{figure*}

\begin{figure*}[t]
  \centering
  \includegraphics[width=\textwidth]{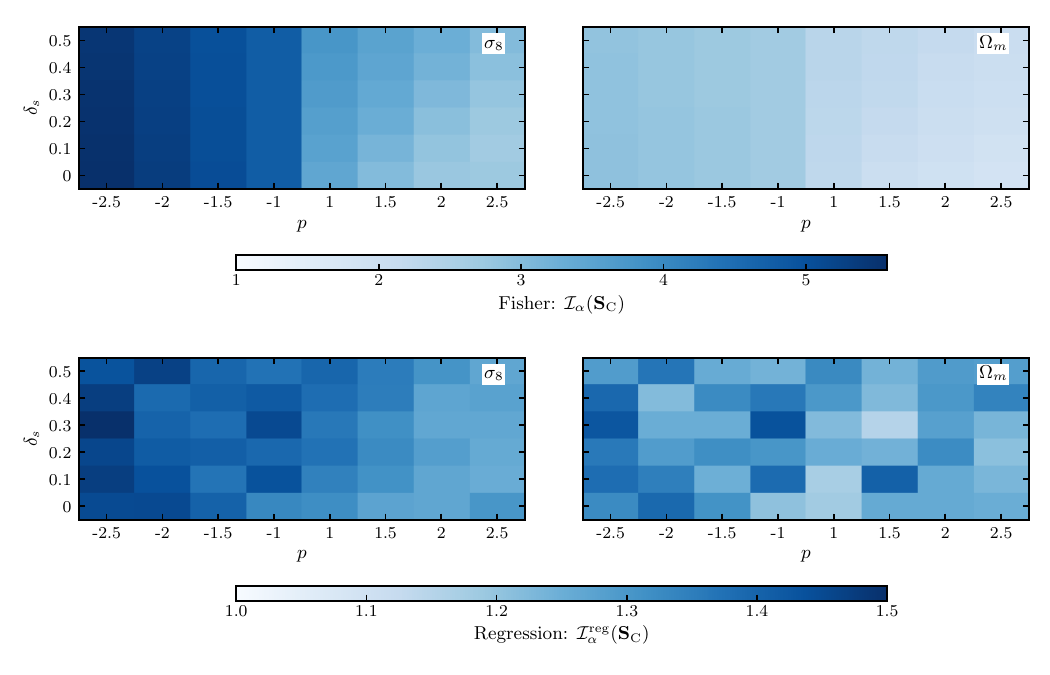}
  \caption{\justifying The Hessian Mark improves on the tested Gaussian~$R=10\,h^{-1}\mathrm{Mpc}$ classical family both locally and across the prior.
Each cell specifies~$(p,\delta_s)$ in \req{eq:classical_mark}, at~$\kmax=0.2\,h\,\mathrm{Mpc}^{-1}$; ratios above one favor the learned mark.
\emph{Top}:~$\mathcal{I}_\alpha(\bS_{\rm C})$, the classical-to-learned marginalized Fisher error ratio at the fiducial cosmology.
\emph{Bottom}:~$\mathcal{I}_\alpha^{\rm reg}(\bS_{\rm C})$, the classical-to-learned RMSE ratio for five-parameter regression, with~$\mathrm{RMSE}_\alpha=\sqrt{\mathrm{MSE}_\alpha}$.
For regression we take the mean of the five RMSE ratios paired by network-initialization seed.
Both rows use the best of the three mark trainings for each parameter.
All cells exceed unity, but the smaller regression ratios show that the local Fisher advantage does not translate into an equal gain across the prior. }
  \label{fig:t_family}
\end{figure*}

The Web Mark already adds information to~$\Pdd$ (\rfig{fig:t_constraints}, right).
For the displayed training, the improvements in~$(\sigma_8,\Omega_m)$ are~$(\mathcal{I}_{\sigma_8},\mathcal{I}_{\Omega_m})=(3.5,2.6)$ at~$\kmax=0.2\,h\,\mathrm{Mpc}^{-1}$ and~$(1.8,1.6)$ at~$\kmax=0.3\,h\,\mathrm{Mpc}^{-1}$.
The Hessian Mark gives larger improvements,~$(\mathcal{I}_{\sigma_8},\mathcal{I}_{\Omega_m})=(6.5,3.0)$ and~$(2.3,1.9)$ at the two scale cuts~$\kmax=0.2-0.3\,h\,\mathrm{Mpc}^{-1}$ respectively.
Across its three trainings, the lower-cut improvements span~$6.4$--$6.5$ for~$\sigma_8$ and~$3.0$--$3.2$ for~$\Omega_m$, so the differences between the mark families exceed this training variation.
The reference classical mark shown in the scale-dependent comparison maximizes the joint five-parameter figure of merit~$\mathrm{FoM}=\sqrt{\det F}$ within this grid at~$\kmax=0.2\,h\,\mathrm{Mpc}^{-1}$, where~$F$ is defined in \req{eq:t_fisher}.
This FoM-selected classical reference provides improvements of~$(2.4,1.7)$ and~$(2.0,1.6)$ in the same scale cuts.
The tighter constraints motivate our focus on the Hessian Mark for the subsequent transfer and regression tests. 
For the common-$8{,}000$-realization comparison with P+B, the Hessian Mark gives gains of approximately~$(5.0,2.6)$ at the lower cut and~$(1.7,1.5)$ at the higher cut (\rfig{fig:t_constraints}, left).

The right panels explain why the gain over~$\Pdd$ is largest at low~$\kmax$.
The Hessian Mark reaches~$\sigma(\sigma_8)\simeq0.0014$ at~$\kmax=0.2\,h\,\mathrm{Mpc}^{-1}$, with little further reduction at~$0.3\,h\,\mathrm{Mpc}^{-1}$.
Thus the added marked bins provide little independent~$\sigma_8$ information after correlations with the existing bins and the other parameters are included.
The nearly constant~$\sigma_8$ error shows that increasing the measured scale cut adds little sensitivity to this parameter.
Because the mark reads the field down to the grid scale, it can transfer information from smaller structures into these large-scale spectra.
We therefore extend P+B to~$\kn=\pi\,128/L\simeq0.402\,h\,\mathrm{Mpc}^{-1}$ while retaining the marked summary at~$\kmax=0.2$ or~$0.3\,h\,\mathrm{Mpc}^{-1}$.
We recompute both forecasts using the same covariance and derivative realizations, five varied parameters, and a Hartlap correction appropriate to each data-vector dimension.
We define the comparison as~$\mathcal{R}_\alpha=\sigma_\alpha[\mathrm{P+B};\kn]/\sigma_\alpha[\bS;\kmax]$, so values above unity favor the marked summary.
Across the three trainings,~$\mathcal{R}_{\Omega_m}=0.98$--$1.02$ and~$\mathcal{R}_{\sigma_8}=1.26$--$1.27$ at~$\kmax=0.2\,h\,\mathrm{Mpc}^{-1}$; at~$\kmax=0.3\,h\,\mathrm{Mpc}^{-1}$, they are~$1.15$--$1.21$ and~$1.30$--$1.31$, respectively.
The large gains over P+B at equal low cuts thus become much smaller when the baseline includes short-wavelength modes, although a residual advantage remains, particularly for~$\sigma_8$.
This comparison supports the interpretation of marking as a compact transfer of small-scale sensitivity into large-scale spectra.

The mark reads the full cubic Fourier grid, whereas the bispectrum cut restricts triangle side-bin centers and uses the existing finite-width bins.
While we restrict the present numerical comparison to P+B, a common input band and a direct higher-order or joint-summary comparison would be needed to attribute the residual gain to particular correlators such as the trispectrum.
In the context of neural field compressions, PatchNet\pcite{Bairagi:2025ytq} combines neural summaries of patches with P+B and reaches constraints close to wavelet scattering\pcite{Valogiannis_2022} on grids of the same resolution.
The Hessian Mark likewise exceeds P+B and reaches comparable error, motivating the possibility that these different compressions approach a common information level. 
Our comparison establishes a gain over P+B on the present simulation setup; the impact of preprocessing and forecast conventions should be systematically treated for a robust comparison of all proposed compressions and it is out of the scope of this work.

\label{sec:t_mnu}
We next test whether the mark extracts information about a parameter absent from its training, the sum of neutrino masses~$M_\nu$.
For this extension we use the cold-dark-matter-plus-baryon field~$\delta_{\rm cb}$ for both the unmarked and marked spectra but we do not retrain the marking network.
At~$M_\nu=0$, this field coincides with the total-matter field used in training; at nonzero mass it retains the gravitational effect of neutrinos on clustering but excludes their direct contribution to the density field.
We apply the same trained filter, gates, and density responses to each cosmology, without retraining or changing their normalization.
For this specific analysis,~$\Pdd$ therefore denotes the power spectrum of~$\delta_{\rm cb}$, and the joint summary remains~$\{\Pdd,\Pdm,\Pmm\}$.
We add~$M_\nu$ to the five cosmological parameters of the fiducial forecast.
At the fiducial value~$M_\nu=0$, we use a one-sided cubic derivative from the Quijote ensembles at~$0$,~$0.1$,~$0.2$, and~$0.4\,\mathrm{eV}$, with~$500$ realizations at each mass\pcite{Villaescusa-Navarro:2019bje}.
The zero-mass derivative reference uses Zel'dovich initial conditions, matching the massive-neutrino ensembles; the covariance uses~$10{,}000$ massless fiducial realizations with second-order Lagrangian initial conditions.
We obtain~$\mathcal{I}_{M_\nu}=1.2$ at~$\kmax=0.2\,h\,\mathrm{Mpc}^{-1}$ and~$1.5$ at~$0.3\,h\,\mathrm{Mpc}^{-1}$. 
Unlike the~$\sigma_8$ ratio, this gain increases with the cut: the added marked bins retain complementary neutrino-mass sensitivity after marginalization over the other five parameters.
Neutrino free-streaming changes the scale dependence of clustering, so the saturation seen for an amplitude parameter need not apply to~$M_\nu$.

Recent work on marked statistics for neutrino-mass constraints provides a comparison with another morphology-based approach\pcite{MartinBarandiaran:2026wwebs}.
With the parameter subset~$(\Omega_m,\sigma_8,M_\nu)$, holding~$\Omega_b$,~$h$, and~$n_s$ fixed, Ref.\pcite{MartinBarandiaran:2026wwebs} combines a classical density mark with a filament mark or a local-fractal-dimension mark, using all six auto- and cross-spectra of the unmarked field and the two marked fields, and reports~$\mathcal{I}_{M_\nu}=2.17$ and~$2.14$, respectively.
At~$\kmax=0.3\,h\,\mathrm{Mpc}^{-1}$ on the same parameter subset, our single mark gives~$\mathcal{I}_{M_\nu}=1.91$ from three spectra.
Our forecast uses a coarser grid ($128^3$ versus~$400^3$ cells) and larger covariance and derivative ensembles ($10{,}000$ versus~$2{,}100$, and~$500$ versus~$50$ per cosmology).
The same study reports~$\mathcal{I}_{M_\nu}=1.93$ for its single classical density mark ($R=10\,h^{-1}\mathrm{Mpc}$,~$p=2$,~$\delta_s=0.25$), compared with our~$1.91$ on that three-parameter subset.
These values are numerically close despite the different resolutions and simulation ensembles; they establish no advantage over that classical reference.
The transfer test instead shows that a mark trained without varying~$M_\nu$ retains sensitivity to it, demonstrating its flexibility in extracting information about parameters beyond its training objective.

A Fisher forecast measures sensitivity at one point of parameter space, while inference needs a summary that separates cosmologies across the whole prior.
Using the same features of \req{eq:t_summary_features}, we train a regression network to predict the five parameters from the marked and unmarked spectra of an unseen set of simulations, as described in \rsec{sec:t_evaluation}.
Unlike the Fisher forecast, this regression also depends on the finite training sample and the inference network.
We employ the same network architecture and training procedure as in Ref.~\cite{Semenzato:2026stt}, with training specifications listed in \rapp{app:round1_training}.
In \rfig{fig:t_family}, we compare the local forecasts and regression errors for a Gaussian family of classical marks with~$R=10\,h^{-1}\mathrm{Mpc}$, at~$\kmax=0.2\,h\,\mathrm{Mpc}^{-1}$.
The upper panel shows marginalized Fisher error ratios, and the lower panel shows RMSE ratios compared to the same classical marks.
At the fiducial cosmology,~$\mathcal{I}_{\sigma_8}(\bS_{\rm C})$ spans~$2.66$--$5.57$ and~$\mathcal{I}_{\Omega_m}(\bS_{\rm C})$ spans~$1.8$--$2.9$ over this family.
At the same scale cut across the prior,~$\mathcal{I}_{\sigma_8}^{\rm reg}(\bS_{\rm C})=1.3$--$1.5$ and~$\mathcal{I}_{\Omega_m}^{\rm reg}(\bS_{\rm C})=1.2$--$1.4$ over the displayed classical family, providing improved generalization across the prior in both parameters.
These values are the mean of the five RMSE ratios paired by network-initialization seed, for the best of the three mark trainings.
The networks are retrained on an extra BSQ subset, disjoint from the simulations used to train the marks.
For this regression architecture and held-out sample, the learned mark has smaller errors than every member of the tested Gaussian~$R=10\,h^{-1}\mathrm{Mpc}$ family at this cut.
At~$\kmax=0.3\,h\,\mathrm{Mpc}^{-1}$ the two converge:~$\mathcal{I}_{\sigma_8}^{\rm reg}(\bS_{\rm C})=0.9$--$1.1$ with mean~$0.98$, and~$\mathcal{I}_{\Omega_m}^{\rm reg}(\bS_{\rm C})=1.0$--$1.2$ with mean~$1.11$, so the clear regression advantage in both parameters is confined to the lower cut.
The local forecast still favors the learned mark across the whole family, with~$\mathcal{I}_{\sigma_8}(\bS_{\rm C})=1.1$--$1.8$ and~$\mathcal{I}_{\Omega_m}(\bS_{\rm C})=1.2$--$1.8$.
While the Fisher forecast employs derivatives around one fiducial cosmology, regression averages prediction errors over a broad prior with a finite training sample and a fixed inference network.
Local derivative sensitivity can therefore improve without extending the same gain in global prediction error.

\FloatBarrier
\section{Conclusions}
\label{sec:t_conclusions}

The non-Gaussian matter distribution contains cosmological information that is difficult to capture with a small set of correlation functions.
Marked spectra offer a way to bring some of that information into two-point inference by changing the relative weights of different regions before measuring their correlations.
Our approach uses local morphology to organize that transformation, then learns its density dependence across a cosmological prior.
The result is a compact summary whose construction remains open to physical inspection and controlled tests.

The Web Mark and Hessian Mark form a progression from fixed environment classes to continuous density--curvature coupling.
The Web Mark assigns a learned density response to each of four environments; the Hessian Mark combines those responses through overlapping weights determined by local curvature.
Keeping the filter, gate weights, and expert responses explicit makes it possible to explore how much information the mark adds and which dependencies supply it.
The ablations favor the full coupling over the tested log-density auto- and cross-spectrum control, a single learned density response, and additive eigenvalue responses at~$\kmax=0.2\,h\,\mathrm{Mpc}^{-1}$.
This progression provides a practical way to increase a mark's flexibility while testing whether each addition earns its complexity.
The Hessian Mark improves constraints beyond the power spectrum and bispectrum at matched output cuts.
At~$\kmax=0.3\,h\,\mathrm{Mpc}^{-1}$, it tightens their marginalized errors by factors of~$1.7$ in~$\sigma_8$ and~$1.5$ in~$\Omega_m$.
Its~$\sigma_8$ error changes little beyond~$\kmax=0.2\,h\,\mathrm{Mpc}^{-1}$, consistent with transferring small-scale sensitivity into large-scale marked spectra.
Extending P+B to the grid Nyquist scale reduces the improvements.
The extended P+B then gives an~$\Omega_m$ error comparable to that of the marked summary at the lower cut, while its~$\sigma_8$ error remains larger by a factor of~$\sim1.26\times$.
The comparison makes the scale transfer explicit without assigning the remaining difference to a particular higher-order correlator.
Regression across the prior keeps an advantage over the whole tested classical family at~$\kmax=0.2\,h\,\mathrm{Mpc}^{-1}$ in both~$\sigma_8$ and~$\Omega_m$; at~$0.3\,h\,\mathrm{Mpc}^{-1}$ the two converge.
The same mark improves neutrino-mass forecasts without retraining, extending the case for learned marks beyond sensitivity at one fiducial cosmology.
The maps retain recognizable void and peak structure at the grid resolution, but visual correspondence alone does not fully identify the origin of the gain.
The modular construction separates these roles without requiring the soft gate weights to act as mutually exclusive web classes.

The present analyses use real-space matter fields at~$z=0$ at limited spatial resolution.
Extending this framework to finer resolutions can test whether resolving wall and filament geometry more finely supplies information beyond the cell-scale curvature measured here, and which morphological features are most relevant for such smaller-scale structure.
More general rotation-equivariant responses could combine density, vector, and tensor features across scales before contracting them into scalar marks, allowing sensitivity to relative alignments as well as eigenvalue shapes.

Application to galaxy surveys requires constructing and testing the mark on sparse, biased tracers in redshift space, with the survey selection applied before estimating its density and curvature inputs\pcite{Marinucci:2024bdq,Ebina:2026qzf}.
Peculiar velocities introduce a preferred line of sight, so the corresponding auto- and cross-spectrum multipoles must be modeled jointly with tracer bias, velocity effects, and stochastic contributions.
Estimating the mark from discrete objects adds fluctuations beyond the shot noise of the original tracer field.
Smoothing and a well-defined treatment of empty cells are necessary before applying the logarithm and spatial derivatives.

Our construction compresses nonlinear clustering information through explicit filters, gates, and density responses whose roles remain interpretable and open to controlled tests.
This modularity provides a concrete route toward marks that combine the information gains of learned transformations with the physical control needed for cosmological inference from surveys.

\section*{Acknowledgements}
In preparing this paper, we used large language models as assistive tools. In particular, LLMs were
used for editing and reviewing the draft, setting up training and numerical analyses and polishing figures. The authors take full responsibility for the content of this paper.
ML acknowledges support by the MIUR Progetti di Ricerca di Rilevante Interesse Nazionale (PRIN) Bando 2022 - grant 20228RMX4A.
AR acknowledges funding from the Italian Ministry of University and Research (MIUR) through the ``Dipartimenti di eccellenza'' project ``Science of the Universe''.

\FloatBarrier
\appendix

\FloatBarrier
\section{Ablations and controls}
\suppressfloats[t]
\label{app:t_ablations}

\begin{table}[h!]
\centering
\caption{\justifying Learned marked ablations.
Entries are~$\mathcal{I}_\alpha$ of \req{eq:t_improvement}, marginalized over the other four parameters, with the same~$\kmax$ applied to every spectrum.
Learned rows report medians over three trainings.}
\label{tab:share_ablations}
\begin{minipage}{0.75\textwidth}
\begin{ruledtabular}
\begin{tabular}{lrrrr}
 & \multicolumn{2}{c}{$\kmax=0.2 h\,\mathrm{Mpc}^{-1}$} & \multicolumn{2}{c}{$\kmax=0.3 h\,\mathrm{Mpc}^{-1}$} \\
Summary &~$\Omega_m$ &~$\sigma_8$ &~$\Omega_m$ &~$\sigma_8$ \\ \hline
Power spectrum + log-density & 1.66 & 2.50 & 1.63 & 2.13 \\
One density expert & 1.61 & 2.45 & 1.56 & 1.97 \\
Density + additive eigenvalues & 1.93 & 3.95 & 1.59 & 2.09 \\
Hard Hessian gates & 2.23 & 3.44 & 1.55 & 2.02 \\
Hessian Mark, learned~$\tau$ & 2.43 & 6.00 & 1.50 & 2.22 \\
Hessian Mark, fixed~$\tau$ & 3.08 & 6.47 & 2.00 & 2.28 \\
\end{tabular}
\end{ruledtabular}
\end{minipage}
\end{table}

The gain of the full mark does not by itself show which parts of its construction are needed.
We therefore test how each component contributes to the final constraints, starting from a fixed transformation and restoring learned structure in stages.
\rtab{tab:share_ablations} compares the resulting summaries at matched scale cuts, with all errors marginalized over the other four cosmological parameters.
Each learned variant is retrained so that its remaining components can adjust to the restriction.
We first test a fixed logarithmic-compression summary, using~$\{\Pdd,P_{gg},P_{\delta g}\}$, where~$g=\ln(1+\delta)$ as in \rsec{sec:background} and~$P_{gg}$ is its auto-spectrum; subtracting its mean changes only the discarded zero mode.
The forecast includes both auto-spectra and their cross-spectrum with their full joint covariance, and has no trained mark.
At~$\kmax=0.2\,h\,\mathrm{Mpc}^{-1}$, this summary provides~$\mathcal{I}_{\sigma_8}=2.50$, well below the full mark's~$6.47$.

We next allow the density response itself to be learned, while removing all gates and eigenvalue inputs.
This one-density-expert mark has~$M=b+u_0(\dfilt)$, with a learned filter and a single learned response to the filtered log-density.
Like all learned rows, it forms the marked field of \req{eq:marked_field} and uses~$\{\Pdd,\Pdm,\Pmm\}$.
Its~$\mathcal{I}_{\sigma_8}$ reaches~$2.45$, close to the fixed log-density control but still below the full gain.
We then restore curvature through separate additive responses,
$M=b+u_0(\dfilt)+\sum_i u_i[\operatorname{sgn}(\lambda_i)\ln(1+|\lambda_i|)]$.
This model reaches~$\mathcal{I}_{\sigma_8}=3.95$, showing the value of eigenvalue dependence on the final constraints.
To test such coupling, we restore the four density experts and use the Hessian eigenvalues to assign their weights.
Hard gates assign each cell to one environment from the eigenvalue signs and give~$\mathcal{I}_{\sigma_8}=3.44$.
Overlapping soft gates let the density response vary continuously with curvature and raise the median~$\mathcal{I}_{\sigma_8}$ to~$6.00$ when the temperature is learned.
Fixing~$\tau=0.6$ in the final design raises the median~$\mathcal{I}_{\sigma_8}$ to~$6.47$, with the training spread reported in \rapp{app:round1_gate_interventions}.
These tests favor the flexible density--curvature coupling at the lower cut.
At~$\kmax=0.3\,h\,\mathrm{Mpc}^{-1}$, the simpler summaries approach the full mark's~$\sigma_8$ constraint as more modes become available.

We further test whether adding explicit curvature responses improves the final fixed-temperature design.
Starting from~$M_{\rm base}=b+\sum_e w_e h_e(\dfilt)$, we separately add either three learned eigenvalue curves,
$\sum_i u_i[\operatorname{sgn}(\lambda_i)\ln(1+|\lambda_i|)]$, or one learned response to the normalized anisotropy
$A=(\lambda_3-\lambda_1)/(\sum_i|\lambda_i|+\epsilon)$, with~$\epsilon=10^{-8}$ in the implementation, increasing the freedom to respond to curvature.
The original gated density terms remain in both models.
The added eigenvalue curves weaken the constraints at both scale cuts, while the anisotropy response gives no consistent gain across the three trainings. 
In further exploratory variants, we also add a fifth density expert controlled by a learned gate in eigenvalue space, sum separate gated experts evaluated at multiple Gaussian smoothing scales, or use a learned gate that combines eigenvalues extracted from fields filtered across different fixed scales. 
These variants give small, parameter-dependent changes relative to their corresponding Hessian baseline, rather than a consistent improvement in constraining power. 
Together with the matched fixed-temperature tests above, they provide no evidence for a robust gain from these additions under the tested training procedures, while not establishing that all useful curvature information has been exhausted.

\subsection{Learned gate distribution}
\label{app:round1_gate_interventions}
We test which dependencies the learned gate fields retain.
For each of the three fixed-temperature seeds, we keep the checkpoint, learned density filter, expert functions and thresholds fixed, and change only how its gates are evaluated. 
The density field and the expert input remain at their original positions.
We compare the original gates with their spatial means in each box, with fixed gate means calibrated on the first~$16$ fiducial boxes, and with gate maps displaced by~$L/2$ along each axis under periodic wrapping. 
We also rescale the eigenvalues by~$s_{\rm fid}/s_{\rm box}$, where~$s_{\rm box}$ is their standard deviation within each realization and~$s_{\rm fid}$ is its mean over the same~$16$ fiducial boxes. 
Per-box statistics are recomputed in every realization, while fiducial constants and the original eigenvalue scale~$s$ remain fixed. 
For each variation, we recompute all three spectra and the five marginalized errors, with the Hartlap correction.
The reference remains the unmarked spectrum on the same boxes and scale cut.

At~$\kmax=0.2\,h\,\mathrm{Mpc}^{-1}$, replacing the gates by their per-box means retains~$\mathcal{I}_{\sigma_8}=6.30$--$6.44$ across trainings, close to~$6.41$--$6.48$ as trained, whereas fixed fiducial means give~$2.04$--$2.13$. 
Thus much of the~$\sigma_8$ sensitivity is carried by variation of the global expert mixture between boxes, which can survive removal of the spatial gate pattern. 
Spatial dependence contributes more to~$\Omega_m$ and~$n_s$: per-box averaging reduces their gains by~$15$--$17\%$ and~$23$--$28\%$, respectively, and displacing the gates also weakens them. 
Rescaling the eigenvalues lowers~$\mathcal{I}_{\sigma_8}$ to~$3.98$--$4.45$. 
At~$\kmax=0.3\,h\,\mathrm{Mpc}^{-1}$, per-box means likewise preserve the~$\sigma_8$ gain while weakening the other constraints. 
These tests distinguish the usefulness of a box-dependent mixture from that of spatial registration; they do not isolate independent information contributions or attribute the full gain to spatial web morphology.

\FloatBarrier
\section{Model specification and training settings}
\label{app:round1_training}
The learned mark schemes consist only of the radial filter, four scalar experts, and the trainable offset. 
Each expert has a~$1\to16\to1$ architecture with GELU activation between the affine layers and zero-initialized output weights and bias. 
The fixed thresholds, temperature, and eigenvalue scale are buffers, not trained parameters.
We form the six independent tensor components by multiplying the FFT of the cell-scale~$g$ by~$k_i k_j$ and inverse transforming.
Numerically we use~$q_i=k_i\Delta x$, with~$\Delta x=L/128$, so the computed tensor is~$(\Delta x)^2H_{ij}$, with this common conversion being absorbed into the fixed eigenvalue scale~$s$. 
All Fourier modes of the cube are retained; for mixed components, we set~$k_i k_j$ to zero on the Nyquist plane of either differentiated axis, preserving the Hermitian symmetry required for a real tensor.
The summary and parameter embedders are used only to train the mark and are discarded when producing its spectra.

\begin{table}[htbp]
\centering
\caption{\justifying Summary of the neural mark and the auxiliary networks used for contrastive training and regression. 
Contrastive embedders use two width-$64$ hidden layers, each followed by LayerNorm and GELU, then a linear output. 
The regression head has hidden widths~$[W,2W,4W,4W,2W,W]$,~$W=3N_k$ with~$N_k$ the number of~$k$ bins, with GELU and dropout~$0.1$ after each hidden affine layer, no LayerNorm, and five linear outputs.}
\label{tab:round1_architecture}
\begin{minipage}{0.8\textwidth}
{\small\setlength{\tabcolsep}{4pt}
\begin{ruledtabular}
\begin{tabular}{ll}
Component & Dimensions or role \\
\hline
Radial filter &~$32$ sigmoid-interpolated knots \\
Four density experts &~$4\times(1\to16\to1)$ \\
Mark offset & One scalar, initialized to~$1$ \\
Deployed mark & Filter + experts + offset \\
Marked embedder &~$94\to64\to64\to16$ \\
Unmarked embedder &~$47\to64\to64\to16$ \\
Parameter projector &~$5\to16$ (linear) \\
Bootstrap optimization & Unmarked embedder + parameter projector \\
Main optimization & Mark + marked embedder + parameter projector \\
\end{tabular}
\end{ruledtabular}
}
\end{minipage}
\end{table}

\begin{table}[htbp]
\centering
\caption{\justifying Optimization settings for bootstrap pretraining of the unmarked reference, mark learning, and subsequent parameter regression. Learning-rate multipliers apply to the base rate; the unmarked embedder is frozen during mark learning.}
\label{tab:round1_optimization}
\begin{minipage}{0.8\textwidth}
{\small\setlength{\tabcolsep}{4pt}
\begin{ruledtabular}
\begin{tabular}{lccc}
Setting & Bootstrap & Mark learning & Regression \\
\hline
Optimizer & AdamW & AdamW & AdamW \\
Base learning rate &~$5\times10^{-4}$ &~$5\times10^{-4}$ &~$10^{-3}$ \\
Weight decay &~$10^{-4}$ &~$10^{-4}$ &~$10^{-4}$ \\
Batch size &~$16$ &~$8$ &~$128$ \\
Maximum epochs &~$550$ &~$700$ &~$1\,000$ \\
Early-stopping patience (epochs) &~$20$ &~$20$ &~$60$ \\
Gradient-clip norm &~$10$ &~$10$ & None \\
Mark learning-rate multiplier & --- &~$2$ & --- \\
Parameter-projector multiplier &~$1$ &~$0.25$ & --- \\
Plateau patience (epochs) &~$8$ &~$8$ &~$8$ \\
Plateau decay factor &~$0.5$ &~$0.5$ &~$0.5$ \\
Minimum learning rate &~$10^{-7}$ &~$10^{-7}$ &~$10^{-7}$ \\
Validation fraction &~$0.15$ &~$0.15$ &~$0.10$ \\
\end{tabular}
\end{ruledtabular}
}
\end{minipage}
\end{table}

\FloatBarrier
\bibliography{bibo}

\end{document}